\documentclass[12pt]{iopart}
\usepackage{iopams}  
\usepackage{graphicx}
\usepackage[rightcaption]{sidecap}

\expandafter\let\csname equation*\endcsname\relax
\expandafter\let\csname endequation*\endcsname\relax
\usepackage{amsmath}
\usepackage{amssymb}
\usepackage{graphicx}
\usepackage{bbold}% bold math
\usepackage[dvipsnames]{xcolor}
\usepackage[pdftex]{hyperref}
\hypersetup{
  linkcolor  = orange!85!black,
  citecolor  = NavyBlue,
  urlcolor   = NavyBlue,
  colorlinks = true,
}
\usepackage{verbatim}

\begin{document}

\title[]{Renormalization group analysis of near-field induced dephasing of optical spin waves in an atomic medium}

\author{S.~Grava$^{1,2}$, Y.~He$^{3}$, S.~Wu$^3$ and D.~E.~Chang$^{1,2}$}

\address{$^{1}$ICFO-Institut de Ciencies Fotoniques, The Barcelona Institute of Science and Technology, 08860 Castelldefels, Barcelona, Spain.}
\address{$^{2}$ICREA-Instituci\'o Catalana de Recerca i Estudis Avan\c{c}ats, 08015 Barcelona, Spain.}
\address{$^3$Department of Physics, State Key Laboratory of Surface Physics and Key Laboratory of Micro and Nano Photonic Structures (Ministry of Education), Fudan University, Shanghai 200433, China.}

\ead{stefano.grava@icfo.eu}

\begin{abstract}
While typical theories of atom-light interactions treat the atomic medium as being smooth, it is well-known that microscopic optical effects driven by atomic granularity, dipole-dipole interactions, and multiple scattering can lead to important effects. Recently, for example, it was experimentally observed that these ingredients can lead to a fundamental, density-dependent dephasing of optical spin waves in a disordered atomic medium.
Here, we go beyond the short-time and dilute limits considered previously, to develop a comprehensive theory of dephasing dynamics for arbitrary times and atomic densities. In particular, we develop a novel, non-perturbative theory based on strong disorder renormalization group, in order to quantitatively predict the dominant role that near-field optical interactions between nearby neighbors has in driving the dephasing process. This theory also enables one to capture the key features of the many-atom dephasing dynamics in terms of an effective single-atom model.
These results should shed light on the limits imposed by near-field interactions on quantum optical phenomena in dense atomic media, and illustrate the promise of strong disorder renormalization group as a method of dealing with complex microscopic optical phenomena in such systems.
\end{abstract}

%\vspace{2pc}
%\noindent{\it Keywords}: Atomic \& molecular optics, YYYYYYYY, ZZZZZZZZZ

%Uncomment for Submitted to journal title message
%\submitto{\NJP}
% Uncomment if a separate title page is required
\maketitle

%-----------------------------------%
\section{\label{sec:Introduction}Introduction}

The interaction of light with atomic ensembles provides the basis for numerous potential applications, such as quantum memories for light \cite{Vernaz-Gris2018,Wang2019}, quantum nonlinear optics with strong photon-photon interactions \cite{Gorniaczyk2014,Tiarks2014,Tiarks2019}, and quantum metrology \cite{Wasilewski2010,Sewell2012,Hosten2016,Chen2014,Cox2016,Lewis-Swan2018}. In order to avoid the complexity associated with the large microscopic number of degrees of freedom, such as the large atom number and their positions, our standard theories for such systems typically favor a macroscopic approach. For example, for atom-light interactions in free space, the Maxwell-Bloch equations (MBE) \cite{Hammerer2010,Bowden1993,Castin1995,Fleischhauer1999,Svidzinsky2015} treat the atoms as a smooth polarizable quantum medium. The MBE have yielded many important insights into the physics that enables the applications above, as well as elucidating performance limitations~\cite{Hammerer2010,Gorshkov2007,Gorshkov2011}. 

Beyond macroscopic phenomena, many microscopic optical effects driven by granularity, dipole-dipole interactions and multiple scattering have been predicted, such as modifications of refractive indices and scattering rates \cite{Morice1995,Ruostekoski1999,Jennewein2016,Jennewein2018,Jenkins2016,Schilder2020}, subradiance \cite{Scully2015,Guerin2016}, and coherent back-scattering \cite{Labeyrie1999, Bidel2002, Aegerter2009}. Besides being of foundational interest, such microscopic effects could also have practical consequences on applications. For example, recently it was experimentally shown in Ref.~\cite{He2021} that such effects lead to a fundamental inhomogeneous broadening of optical transitions in an ensemble. This  manifests itself as an additional dephasing on top of spontaneous emission decay for optical spin waves, with a rate that is exponential at early times and is directly proportional to atomic density. It was argued that this initial dephasing arises from the strong near-field interaction of a small fraction of particularly close nearest neighbors, quantitatively reproducing the experimental results. Separately, though, one might wonder what governs the apparently non-exponential behavior at later times, or what occurs at very high densities, when many atoms sit within a wavelength of each other and experience strong near-field interactions. We also note that near-field interactions have been recognized to play key roles in other collective behavior, ranging from the modification of superradiance in small systems \cite{Gross1982} to late-time subradiance \cite{Cipris2021} in extended systems. Beyond exact numerics, however, development of effective theories generally remains a challenge in many-atom disordered systems.

\begin{figure}
 \centering
 \includegraphics[trim={0 0 1.7cm 0},height=2.7cm]{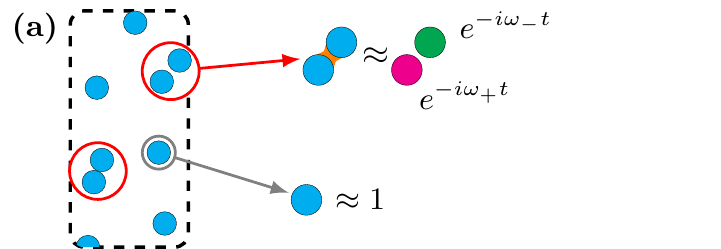}
 \includegraphics[height=2.7cm]{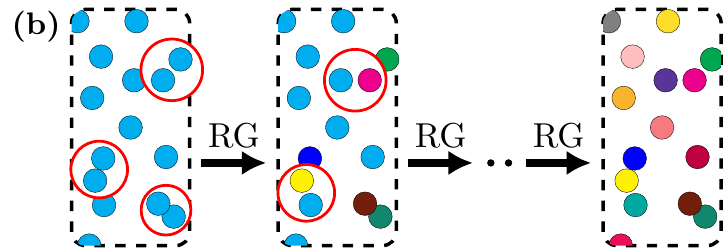}
 \caption{(a) A pair-wise approach to the many-atom optical dynamics. 
 As derived in Ref.~\cite{He2021}, in a dilute atomic medium, a small fraction of pairs of atoms (red circles) are separated by a distance much smaller than a wavelength, and thus interact strongly via their near fields. These atoms can be replaced with a non-interacting, dynamically equivalent pair with new effective frequencies (indicated by different colors), which evolve with the phase $e^{-i\omega_{\pm}t}$. Single isolated atoms instead will not significantly contribute to time evolution. (b) Representation of the RG scheme~\cite{Andreoli2021}. Each step is characterized by identifying the most strongly interacting pairs and replacing them with two new atoms with different frequencies, which do not interact anymore through the near field. Unlike in Fig.~\ref{fig:RG_scheme} (a), one can continue this process (including the renormalization of atoms previously renormalized) until all near-field interactions have been eliminated. The overall system at the end is equivalent to an inhomogeneously broadened ensemble of spectral distribution $P(\omega)$, plotted in Fig.~\ref{fig:RG_Pw}.}
 \label{fig:RG_scheme}
\end{figure}

Here, we provide a comprehensive theoretical picture of the spin-wave dephasing phenomenon by applying a non-perturbative technique based on strong disorder renormalization group (RG), which is a powerful method to elucidate the physics in diverse disordered condensed matter systems \cite{Levitov1990,Fisher1994,Damle2000,Motrunich2000,Refael2004,Igloi2005,Vosk2013,Refael2013} and has also recently been applied to atom-light interactions~\cite{Andreoli2021}. As in the short-time theory of dephasing, one key idea underlying this approach is that for highly disordered atomic media, strong near-field interactions between particularly close nearest neighbors allow such pairs to be approximately diagonalized first. The resulting dynamics is equivalent to replacing the pair with two, new \textit{effective} atoms with renormalized frequencies, as illustrated in Fig.~\ref{fig:RG_scheme} (a). The RG theory goes significantly beyond this, however, by realizing that nearby, strongly interacting pairs (including atoms previously renormalized) can continue to be identified and diagonalized, i.e. the many-atom system interacting via the near field can be thought of and diagonalized in terms of an extended hierarchy of strongly interacting pairs~(Fig.~\ref{fig:RG_scheme} (b)). The final result is that the original system is optically equivalent to an inhomogeneously broadened medium with a well-defined distribution of resonance frequencies $P(\omega)$, and with the strong near-field interactions effectively removed. This approach was recently used to predict that a disordered atomic medium has a limiting value of maximum refractive index, regardless of its physical density~\cite{Andreoli2021}. Here, we show that RG not only works to capture the stationary optical response of a dense gas, but also to capture the above mentioned time-dependent dephasing dynamics of spin waves, in a simple and non-perturbative way. The validity of the RG approach is quantitatively verified by comparison with full, microscopic coupled-dipole simulations of large ($N\sim 10^4$) atomic ensembles.

%----------------------------------------------------%
%---------------------------------------------------%

The remainder of the paper is structured as follows. 
In Sec.~\ref{sec:microscopic_model} we briefly review the microscopic theoretical description of photon-mediated dipole-dipole interactions, which accounts for atomic positions, near-field interactions, and multiple scattering of light, and which serves as the basis for the microscopic simulations of spin-wave dynamics. In Sec.~\ref{sec:RG} we describe the RG approach, which enables one to predict a universal inhomogeneous broadening function for a disordered medium. From here, we then formulate a simple, approximate, \textit{single-atom} model for the spin-wave dephasing dynamics. In Sec.~\ref{sec:spin_wave_dephasing}, we present detailed numerical simulations of the spin wave dynamics from dilute to high-density media, which show both the initial exponential dephasing and non-exponential behavior at later times. We also compare these results with the RG approach, which exhibits good quantitative agreement in all regimes. We conclude and provide an outlook in Sec.~\ref{sec:conclusions}.

\section{\label{sec:microscopic_model}Microscopic model of atom-light interaction dynamics}

We consider a minimal model consisting of $N$ identical 2-level atoms at fixed, random positions $\left\{\mathbf{r}_{i}\right\}_{i=1,\dots,N}$ that are uniformly distributed within a spherical cloud. The ground and excited states $|g_{j}\rangle$ and $|e_j\rangle$ have an electric dipole transition characterized by resonance frequency $\omega_{0}=ck_0$ and wavelength $\lambda_0=2\pi/k_0$, and a single-atom excited-state spontaneous emission rate given by $\Gamma_0$. We also define a dimensionless density in terms of the number of atoms per cubic wavelength $\eta=\lambda_{0}^3N/V$, where $V=\frac{4}{3}\pi R^3$ is the volume of the ensemble and $R$ its radius.

Within the standard assumptions in quantum optics~(dipole, rotating wave and Markov approximations), the effects of photon-mediated dipole-dipole interactions, multiple scattering, and wave interference in spontaneous emission are captured by an effective atomic Hamiltonian \cite{Agarwal1970,Gross1982,Asenjo-Garcia2017}: 
%--------------------------------%
\begin{equation}
   H=-\frac{3\pi\Gamma_0}{k_0}
   \sum_{j\ell}
	\mathbf{d}^{*}
	\cdot
	\mathbf{G}_{0}(\mathbf{r}_{j}-\mathbf{r}_{\ell},\omega_{0})
	\cdot
	\mathbf{d} \ \sigma_{eg}^{j}\sigma_{ge}^{\ell}.
	\label{eq:CDM}
\end{equation}
%--------------------------------%
Physically, the photon-mediated interactions between atoms are described by the free-space Green's tensor $\mathbf{G}_{0}(\mathbf{r},\omega_0)$, the fundamental solution of the wave equation \cite{Novotny2006}. The Green's tensor  characterizes how a photon emitted by an atom at $\mathbf{r}_j$, via the action of the atomic lowering operator $\sigma_{ge}^{j}=|g_{j}\rangle\langle e_j|$, propagates to a second atom at $\mathbf{r}_i$, travelling a distance $r_{j\ell}=|\textbf{r}_{j\ell}|=|\mathbf{r}_{j}-\mathbf{r}_{\ell}|$. Moreover, as $\mathbf{d}$ here represents the orientation of the atomic dipole matrix element, which we fix to be along $\hat{\mathbf{x}}$, we conveniently define $\theta_{j\ell}$ as its angle with respect to $\mathbf{r}_{j\ell}$. Doing so, the Hamiltonian $H=\sum_{j\ell} H_{j\ell}\sigma_{eg}^{j}\sigma_{ge}^{\ell}$ is characterized by the coefficients
%---------------%
\begin{equation}
    H_{j\ell}=-\frac{3\Gamma_0}{4}e^{ik_0r_{j\ell}}
    \left[
        \frac{3\cos^2\theta_{j\ell}-1}{(k_0r_{j\ell})^3} 
        -i\frac{3\cos^2\theta_{j\ell}-1}{(k_0r_{j\ell})^2}
        -\frac{\cos^2\theta_{j\ell}-1}{k_0r_{j\ell}} 
    \right].
    \label{eq:H_r_costheta}
\end{equation}
%---------------%
The Hamiltonian is symmetric and non-Hermitian as a consequence of $\mathbf{G}_{0}$ enforcing reciprocity and being a complex quantity, including both coherent and dissipative interactions. 

Important to later discussions, the real part of the Green's function~(describing coherent interactions) contains a $\sim 1/r^3$ near-field component, which dominates at small inter-atomic distances ($k_0 r\!<\!1$). This near-field interaction explicitly reads: 
%---------------%
\begin{equation}
    H_{j\ell}^{\text{near}}=-\frac{3\Gamma_0}{4(k_0r_{j\ell})^3}\left(3\cos^2\theta_{j\ell}-1\right).
    \label{eq:H_near}
\end{equation}
%---------------%
The dissipative part instead describes collective spontaneous emission as arising from wave interference of the emitted light, while in the limit of a single atom, predicts the known spontaneous emission rate of $\Gamma_0=\omega_{0}^3d_{eg}^2/3\pi\hbar\varepsilon_0c^3$, with $d_{ge}$ being the amplitude of the dipole matrix element for the atomic transition. Generally, the presence of dissipation requires a master equation treatment~\cite{Gross1982,Dung2002,Buhmann2004,Buhmann2012}, but the non-Hermitian Hamiltonian \eqref{eq:CDM} is sufficient to describe the single-excitation regime of interest in our work, which generally reduces to solving a set of classical coupled dipole equations of motion \cite{Morice1995,Novotny2006,Svidzinsky2015,Zhu2016,Asenjo-Garcia2017,Skipetrov2014,Zhu2016,Araujo2016,Bettles2016,Bromley2016}.

We will specifically be interested in applying the Hamiltonian \eqref{eq:H_r_costheta} above to investigate the dynamics of a single-excitation ``timed Dicke state'' or spin wave, defined as
\begin{equation}
   |\mathbf{k}\rangle=\frac{1}{\sqrt{N}}\sum_{j} e^{i\mathbf{k}\cdot\mathbf{r}_j}|e_j\rangle.
   \label{eq:spin_wave}
\end{equation}
These collective states with well-defined wavevector constitute a natural basis to describe light-matter excitations. For example, phase-matched spin waves with $|\mathbf{k}| \approx k_0 = \omega_0/c$ are naturally and easily excited by an incoming resonant short pulse. By reciprocity, it is well known that they also efficiently emit into a narrow, well-defined direction centered around $\mathbf{k}$ \cite{Scully2006}, with a collectively enhanced rate \cite{Scully2009,Friedberg2010,Bienaime2011,Svidzinsky2015,Araujo2016,Zhu2016,Guerin2017,Bromley2016,He2020}, $\Gamma_{|\mathbf{k}|\sim k_0}^{\text{coll}}/\Gamma_0=1+\overline{\text{OD}}/4$, which linearly scales with the average optical depth of the medium. This narrow emission occurs due to constructive interference of the emitting atoms along the $\mathbf{k}$ direction, and forms the basis of collective enhancement at the heart of efficient atom-light interfaces~\cite{Hammerer2010,Fleischhauer2005} and the applications mentioned in the introduction. This behavior can be equally derived by microscopic theories \cite{He2020} or by the macroscopic MBE \cite{Scully2009,Friedberg2010,Svidzinsky2015}.

It should be noted that phase-matched spin wave excitations undergo non-trivial macroscopic spatio-temporal propagation dynamics \cite{Cottier2018,DoEspiritoSanto2020}. This makes it challenging to quantify the magnitude and effects of microscopic dephasing, due to the difficulty in defining an ideal time-evolving reference state to compare to, if dephasing could hypothetically be eliminated. An elegant, robust solution to this problem was proposed in Ref.~\cite{He2021}, with experimental realization based on a series of time-domain spin-wave control techniques~\cite{He2020,He2020a}. In particular, one can excite mismatched spin-wave excitations $|\mathbf{k}\rangle$, characterized by $|\mathbf{k}| \neq k_0$, which neither couple to light efficiently nor have any preferred emission or propagation direction. Due to the lack of any direction in which the emitted field~(averaged over random configurations) interferes, the average initial spontaneous emission rate of this state reduces to the single-atom value $\Gamma_{|\mathbf{k}|\neq k_0}^{\text{coll}}=\Gamma_0$ \cite{Scully2015}. Since this state does not exhibit any background macroscopic dynamics, $|\mathbf{k}\rangle$ itself serves as a reference to compare against the actual time-evolved state.

%-----------------------%
\begin{figure*}
    \centering
    \begin{minipage}[t]{0.714\linewidth}
    \includegraphics[width=\linewidth]{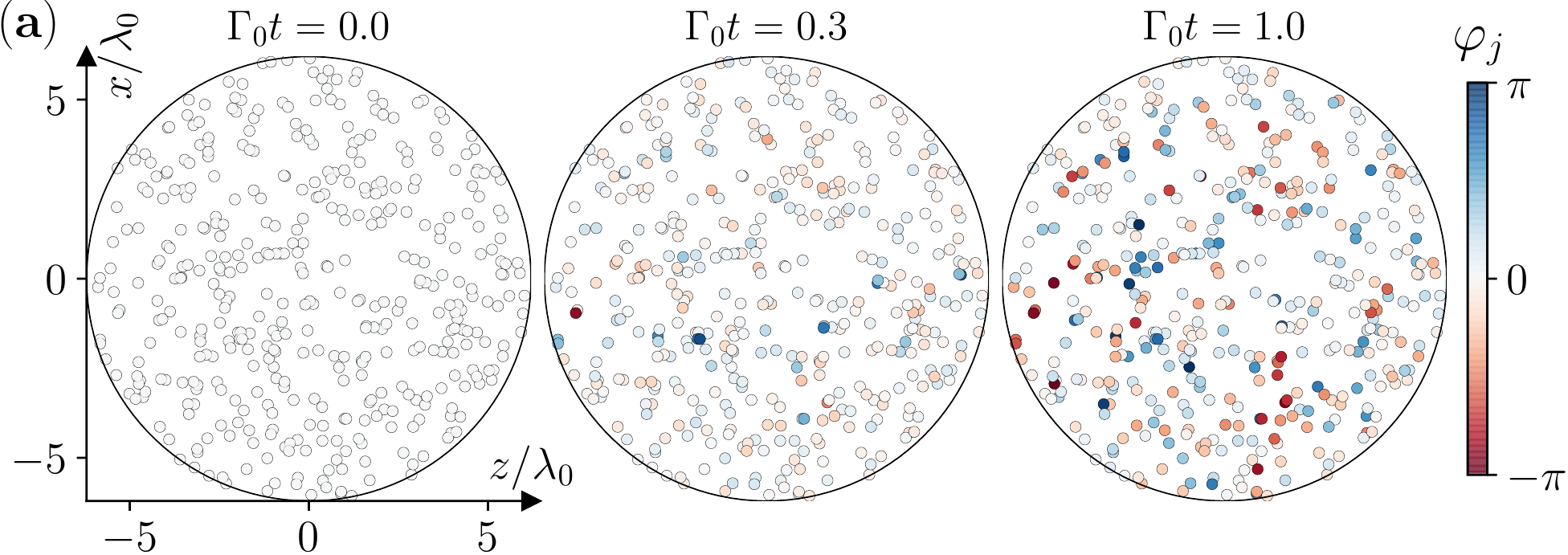}
    \end{minipage}
    \begin{minipage}[t]{0.236\linewidth}
    \includegraphics[width=\linewidth]{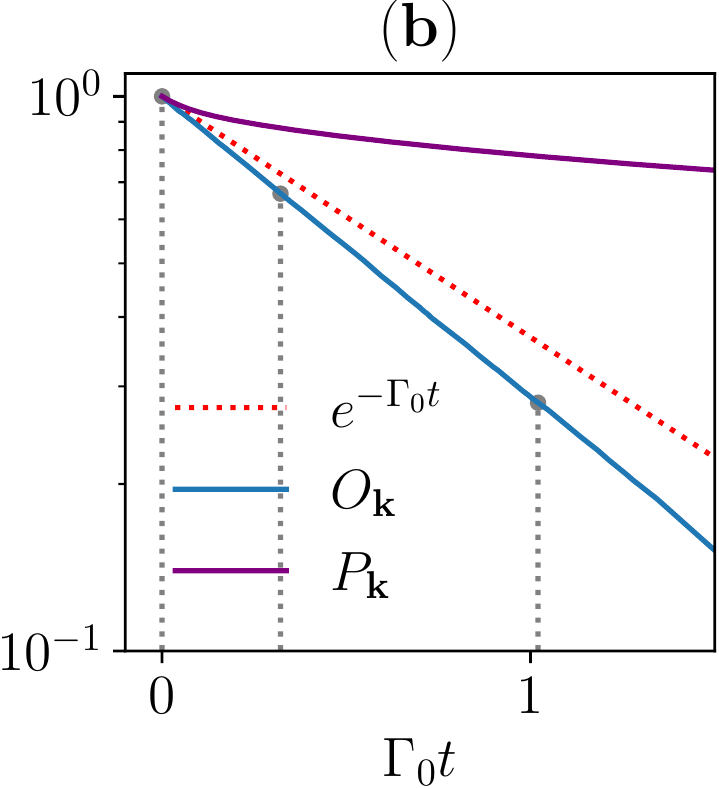}
    \end{minipage}
    \caption{(a) Snaphots of the time evolution of a mismatched spin-wave. We initially prepare a mismatched spin-wave $|\mathbf{k}\rangle$ ($|\mathbf{k}|=6.0k_0$) in a particular configuration $\{\mathbf{r}_j\}$ of a disordered gas at density $\eta=10$, $N=10^4$ atoms and radius $R/\lambda_0\sim 6$. The state is then let to evolve under the dipole-dipole Hamiltonian \eqref{eq:CDM} and the projection over the initial state is computed, $\langle\mathbf{k}|\mathbf{k}(t)\rangle=\sum_{j}|f_j|e^{i\varphi_j}$, extracting the time dependent amplitudes and phases in the single atom basis. To create the snapshots above we consider all the atoms contained in a slice of size $\Delta y=R/35$, plot their position along the $xz$ plane, and color them according to their accumulated phase in time evolution $\varphi_j$, as a consequence of the interaction, for different times. The global effect of this dephasing is represented in (b) where we plot the global overlap $O_{\mathbf{k}}=|\langle\mathbf{k}|\mathbf{k}(t)\rangle|^2$ (blue curve) and population $P_{\mathbf{k}}=\langle\mathbf{k}(t)|\mathbf{k}(t)\rangle$ (purple) of the same mismatched spin-wave. Quantifying the deviation of the overlap with respect to the predicted decay $e^{-\Gamma_0t}$ (red dotted line) is the main purpose of this work.}
    \label{fig:visual_dephasing}
\end{figure*}
%-----------------------%

Specifically, we will be interested in the time evolution under the dipole-dipole Hamiltonian \eqref{eq:CDM} as given by $|\mathbf{k}(t)\rangle=e^{-iHt}|\mathbf{k}\rangle$.
Since the initial state contains only a single excitation, as argued above, the dynamics can be efficiently evaluated numerically, with the resulting equations of motion equivalent to classical coupled dipole equations. From the time-dependent state, we can construct two quantities of interest, 
\begin{equation}
 \begin{aligned}
   P_\mathbf{k}(t)&=\langle\mathbf{k}(t)|\mathbf{k}(t)\rangle\\
   O_\mathbf{k}(t)&=|\langle\mathbf{k}|\mathbf{k}(t)\rangle|^2.
 \end{aligned}
 \label{eq:Pop&Overlap}
\end{equation}
The first quantity $P_\mathbf{k}(t)\leq 1$ monotonically decreases and gives the total remaining excited state population at any time $t$, with the rest having been irreversibly lost due to~(collective) spontaneous emission. The second quantity, $O_\mathbf{k}(t)$, on the other hand, quantifies the overlap with the initial spin wave, and thus describes the survival of the spin-wave order. Importantly, it was shown to be a measurable quantity in the experiments of Ref.~\cite{He2021}.

While we will present a more systematic analysis in Sec.~\ref{sec:spin_wave_dephasing}, we provide a visual example of the physics encoded in $P_\mathbf{k}(t)$ and $O_\mathbf{k}(t)$ in Fig.~\ref{fig:visual_dephasing}. In particular, we simulate the dynamics of an initial spin wave $|\mathbf{k}\rangle$, with $|\mathbf{k}|=6k_0$, in a particular configuration $\{\mathbf{r}_j\}$ of a disordered gas of $N=10^4$ atoms with density $\eta=10$, and in a spherical volume of radius $R/\lambda_0\approx 6$. The evolved state $|\mathbf{k}(t)\rangle$ under Eq.~(\ref{eq:CDM}) is calculated for several specific times $t$. In Fig.~\ref{fig:visual_dephasing} (a) we consider all the atoms contained within a slice $\Delta y=R/35$ of the center of the cloud, and plot their positions in the $x$-$z$ plane. The colors represent the accumulated phase of each atom relative to its initial value $e^{i\mathbf{k}\cdot\mathbf{r}_j}$, with a strong dephasing evident at time $\Gamma_0 t=1.0$. In Fig.~\ref{fig:visual_dephasing} (b), we then plot $P_\mathbf{k}(t)$~(purple) and $O_\mathbf{k}(t)$~(blue) as calculated for the entire ensemble, along with the single-atom spontaneous emission $e^{-\Gamma_0 t}$ for reference~(dashed red). It can be seen that while the initial decay of the total excited population $P_\mathbf{k}(t)$ occurs at a rate $\sim\Gamma_0$~(confirming the absence of collective enhancement) before slowing down, the spin wave survival ratio $O_\mathbf{k}(t)$ decays significantly faster than $\Gamma_0$, due to the dephasing illustrated in Fig.~\ref{fig:visual_dephasing} (a).

%While here we will focus on the case of mismatched spin-waves, we conclude this section acknowledging that other authors studied the effect of microscopic near field interaction in other states \cite{Cipris2020VanClouds} (late-time subradiant), and that corrections the prediction $\Gamma_{|\mathbf{k}|\neq k_0}^{\text{coll}}=\Gamma_0$ is generically expected when collective macroscopic emission is suppressed, as a consequence granularity of the atomic distribution and dipole-dipole interaction \cite{Gross1982Superradiance:Emission}. While these microscopic details makes the problem quite complex in general, in the next section we provide a simple non-perturbative solution of the dynamics.

\section{\label{sec:RG}A Renormalization group approach}

While the microscopic model (Eq.~\eqref{eq:CDM}) can be numerically solved for moderate atom number, its complexity scales directly with the number of atoms $N$, and with the number of configurations needed to obtain disorder-averaged results. Furthermore, the exact numerics does not directly elucidate the underlying physics. Motivated by that, here we introduce a simple model, based on a strong-disorder RG approach, which clearly identifies the role that near-field interactions have on the dynamics, and which allows the effects of such interactions to be captured by a simple \textit{effective, single-atom} theory.

In a previous work \cite{He2021}, we derived a theory to understand the short-time dephasing rate of a spin wave for dilute ensembles with densities $\eta\lesssim 1$. We then found that the dephasing is primarily attributable to a small fraction of atomic pairs separated by a distance much smaller than the optical wavelength, $k_0 r\!<\!1$ (highlighted in Fig.~\ref{fig:RG_scheme} (a)), which strongly interact via their near fields (Eq.~\eqref{eq:H_near}).  The $\sim 1/r^3$ scaling of the near field implies~(in three dimensions) that the presence of other atoms is just a weak perturbation on top of the strong pairwise interaction, such that the pair can be separately and approximately diagonalized. In the single-excitation manifold, diagonalizing the near-field interaction \eqref{eq:H_near} of a pair yields symmetric and  anti-symmetric eigenstates, $|\pm\rangle=(|eg\rangle \pm |ge\rangle)/\sqrt{2}$, which experience opposite frequency shifts $\omega_{\pm}=\pm\frac{3\Gamma_0}{4k_0^3 r^3}(1-3\cos^2\theta)$ relative to the bare atomic transition frequency.

The time evolution for the two-body problem can now be studied in terms of its normal modes. Concretely, in the single atom rotating frame $e^{-i\omega_0t}$, an initially prepared two-body spin-wave $|\mathbf{k}\rangle$, will evolve as $\langle\mathbf{k}|\mathbf{k}(t)\rangle=e^{-i\omega_{+}t}|c^{+}_{\mathbf{k}}|^2 + e^{-i\omega_{-}t}|c^{-}_{\mathbf{k}}|^2$, having defined the projections $c^{\pm}_{\mathbf{k}}=\langle\mathbf{k}|\pm\rangle$.
Although the magnitude of $\mathbf{k}$ might be constrained in an experiment, we can take the conceptual limit where $k\rightarrow\infty$, or infinite mismatch. This implies that the phase $e^{i\mathbf{k}\cdot\mathbf{r}_j}$ of each excited atom is effectively random~(being infinitely sensitive to the specific atomic position), which implies that the actual spin wave should have on average equal overlap with the $\pm$ eigenstates, i.e. that $|c^{\pm}_{\mathbf{k}}|^2\rightarrow 1/2$. The dynamics can therefore be equivalently modeled by replacing the two original atoms with two new atoms of new resonance frequencies $\omega_{\pm}$, that now do not interact anymore through the near field, but evolve with their ``free'' inhomogeneous phases $e^{-i\omega_{\pm}t}$ (see Fig.~\ref{fig:RG_scheme} (a)).
%\textcolor{orange}{I feel we shall follow up on this approximation somewhere later in the paper, by pointing out that the directional dependence of the dipolar interaction ensures that the tails of $P(\omega)$ is symmetric in a random gas, even if the +- modes are excited in an imbalanced fashion -- This is perhaps obvious at the first step of this RG process, I think, but is it probably generalizable to the whole RG process? More generally, I feel a gap of reasoning is here related to an effectively invariant interaction matrix element during RG. The related issue seems to be resolved in [41], not relying on $k\rightarrow \infty$ approximation.  Would it be possible to address this gap similarly here?  Well, I also see that maybe by just leaving a gap here could call curiosity from readers, and that the analysis are supported numerically anyway. So I do not hold a strong opinion on this.  Nonetheless,  I feel it might be nice to even just briefly discuss on this issue and cite similar discussions in \cite{Andreoli2021MaximumMedium}.}

In Ref.~\cite{He2021}, the short-time dynamics of a many-atom, dilute ensemble were approximated by identifying a small fraction of close-by atomic pairs that evolved with the phases $e^{-i\omega_{\pm}t}$ as argued above, while the remaining atoms were assumed to undergo no evolution. By taking the position dependence of the function $\omega_{\pm}(\mathbf{r})$ and combining with the known distribution function of separations $\mathbf{r}$ of nearest neighbors in a random ensemble, an initial decay of the spin wave survival order was predicted and measured to be exponential, $|O_{\mathbf{k}}(\delta t)|^2=e^{-\Gamma_\mathbf{k}\delta t}$, with a density-dependent rate $\Gamma_\mathbf{k}/\Gamma_0=1+\xi\eta$, where $\xi=1/6\pi\sqrt{3}$ for two-level atoms. 

Now, in a dense ensemble (Fig.~\ref{fig:RG_scheme} (b)) or at longer evolution times, the basic picture of replacing close-by atomic pairs with new effective atoms of renormalized frequencies does not change. However, a key realization is that after an atom has been renormalized, it can still see another atom close by with which it can strongly interact (again highlighted by red circles in Fig.~\ref{fig:RG_scheme} (b)). This allows yet another renormalization step to take place, which now will involve the diagonalization of a pair of atoms with the effective frequencies previously obtained. Whereas only a small fraction of atoms dictates the initial decay in a dilute ensemble, here, we must specify a general procedure valid for any density, to repeatedly and hierarchically identify the single most strongly interacting pair~(including the possibility that the pair contains already renormalized atoms) and replace them with two new effective atoms. 

Having anticipated that RG can involve the renormalization of atoms that have already been renormalized in previous steps, here, we consider the more general case of two inhomogeneous atoms of general resonance frequencies of  $\omega_i$ and $\omega_j$, interacting via the near field, as described by the two-body Hamiltonian in the single excitation sector,
\begin{equation}
	 H_{j\ell}^{2b}=\langle\omega_{j\ell}\rangle\mathbb{1}+
	 \begin{pmatrix}
	 	\delta\omega_{j\ell}	&	H_{j\ell}^{\text{near}}\\
	 	H_{j\ell}^{\text{near}}	&	-\delta\omega_{j\ell}
	 \end{pmatrix}.
 \label{eq:2body_H}
\end{equation} 
For convenience, we have defined $\langle\omega_{j\ell}\rangle=(\omega_{j}+\omega_{\ell})/2$, $\delta\omega_{j\ell}=(\omega_{j}-\omega_{\ell})/2$. To quantify the strength of the interaction, we define the ratio between the off-diagonal and the diagonal elements, $\mathcal{K}_{j\ell}=\mathcal{L}_{j\ell}|H_{j\ell}^{\text{near}}|/(|\delta\omega_{j\ell}|+1)$, where the matrix $\mathcal{L}$ keeps the information of whether a pair of atoms has already been renormalized ($\mathcal{L}_{j\ell}=0$) or not ($\mathcal{L}_{j\ell}=1$). This prevents a pair of renormalized atoms from being renormalized between themselves multiple times~(although each atom from the pair can be renormalized with other atoms). Intuitively, a large value of $\mathcal{K}_{j\ell}$ (which requires $\mathcal{L}_{j\ell}=1$) means that the strength of the interaction is able to further split the original frequency difference $\delta\omega_{j\ell}$. Thus, the most strongly interacting pair is identified as that with the largest value of $\mathcal{K}_{j\ell}$ (red circles in Fig.~\ref{fig:RG_scheme}).
Once identified, the full diagonalization of \eqref{eq:2body_H} gives two eigenvalues, $\omega_{\pm}=\langle\omega_{j\ell}\rangle+\sqrt{\delta\omega_{j\ell}^2+\left(H_{j\ell}^{\text{near}} \right)^2}$. The pair can be therefore replaced by an approximately equivalent one, made of two atoms with the new resonance frequencies $\omega_{\pm}$ and that do not interact anymore through the near field (setting $\mathcal{L}_{j\ell}=0$).

\begin{figure}
 \centering
  \includegraphics[width=0.47\textwidth]{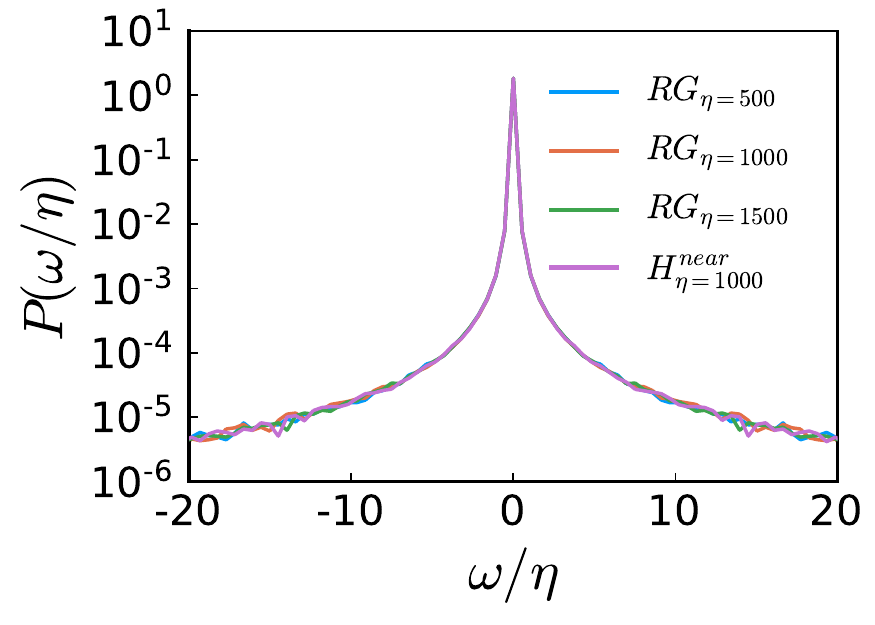}
    \caption{Universal probability distribution of normalized frequencies $\omega/\eta$. We apply the RG approach to spherical samples of $N=2500$ atoms at dimensionless densities $\eta=500,1000,1500$ to extract the new effective frequencies $\left\{\omega_i\right\}$, sampling approximately $\sim10^3$ different configurations at each density to build the probability distribution $P(\omega/\eta)$. The purple solid line instead corresponds to the exact numerical spectrum of the near-field interaction Hamiltonian. This is obtained considering an ensemble of atoms (same parameters as before) and diagonalizing $H^{\text{near}}$ (defined in Eq.~\eqref{eq:H_near}) to get the eigenfrequency statistics.}
 \label{fig:RG_Pw}
\end{figure}

Repeated application of this algorithm, which constitutes the renormalization group (RG) flow, continues until all near-field interactions have been removed and the atoms have been assigned the new effective frequencies  $\left\{\omega_i\right\}$. When the RG scheme is applied to multiple realizations of randomly distributed ensembles of atoms, we can build up the probability distribution $P(\omega)$ of the effective frequencies, which we illustrate in Fig.~\ref{fig:RG_Pw}. 

As the near-field interaction only depends on distance through the dimensionless parameter $(k_0 r)^{-3}$, the distribution when rescaled by density, $P(\omega/\eta)$, should be a universal function for a sufficiently large number of atoms and sufficiently large geometry, where boundary effects are negligible. We can directly confirm this numerically in Fig.~\ref{fig:RG_Pw}, where we plot $P(\omega/\eta)$ obtained from RG for various densities.

%------------------------------------------%
%---------------------------------------------%

The high-frequency tails of $P(\omega)$ correspond to the most strongly interacting pairs, which are renormalized at the beginning of the flow. This perfectly matches the simpler theory presented in \cite{He2021} for dilute atoms, based on the probability distribution of nearest neighbors, where it was found that asymptotically
\begin{equation}
    P(\omega)\sim_{\pm\infty}\xi\eta\frac{1}{2\pi\omega^2}
    \label{eq:Pw_tails}
\end{equation}
The central part of the distribution $P(\omega)$ instead consists of atoms that have been renormalized multiple times.
In this sense, the presented RG scheme and the resulting probability distribution is the correct way to capture the near field induced inhomogeneous broadening of the medium and the induced dephasing rate of spin-waves in dense media, as we are going to discuss.

Formally, the RG procedure amounts to approximately diagonalizing the near-field part of the Hamiltonian~\eqref{eq:CDM} (an $N\!\times\!N$ matrix), by repeatedly identifying and diagonalizing a dominant interacting pair of atoms (a $2 \times 2$ block). We can quantify the error by comparing the resulting frequency distribution $P(\omega)$ obtained by RG, with the eigenvalue distribution obtained by exact, numerical diagonalization of the real, symmetric $N \times N$ Hamiltonian $H^{\text{near}}$ defined by only considering the near field interaction (Eq.\eqref{eq:H_near}). It can be seen in Fig.~\ref{fig:RG_Pw} that the two are essentially indistinguishable, which validates the RG scheme.

We now discuss how the RG results can be applied to predict the time evolution of a spin wave in an effective single-atom picture, focusing on the spin-wave survival ratio $O_{\bf k}(t)$. In particular, while the coherence of a single, isolated atom (without the rotating frame) is expected to evolve as $\langle\sigma_{ge} (t)\rangle = \langle\sigma_{ge} (0)\rangle e^{-i(\omega_0 -i\Gamma_0/2)t}$, the distribution in resonance frequencies of the new effective medium will result in an uncertainty of accumulated phase in time
%----------------------------%
\begin{equation}
	O_{\textsc{rg}}(t)=\left|\int d\omega P(\omega)e^{i\omega t}\right|^2
	e^{-\Gamma_0t},
	\label{eq:overlap_RG}
\end{equation}
%---------------------------%
thus introducing microscopically-driven dephasing due to inhomogeneous broadening. We note that the simplicity of Eq.~\eqref{eq:overlap_RG} arises in part due to our choice of a phase-mismatched spin wave as the initial state. In particular, although RG removes the strong near-field interactions, the atoms can still interact and radiate via the far field, which as we described earlier can in general result in non-trivial spatio-temporal dynamics even within the MBE. The use of a phase-mismatched spin wave suppresses the macroscopic dynamics, leaving the predicted spin wave survival ratio to depend on the microscopic quantity $P(\omega)$ alone.

\section{\label{sec:spin_wave_dephasing}Spin-wave dephasing}
%----------------------------------------------------------%
\begin{figure*}
    \centering
    \includegraphics[width=\linewidth]{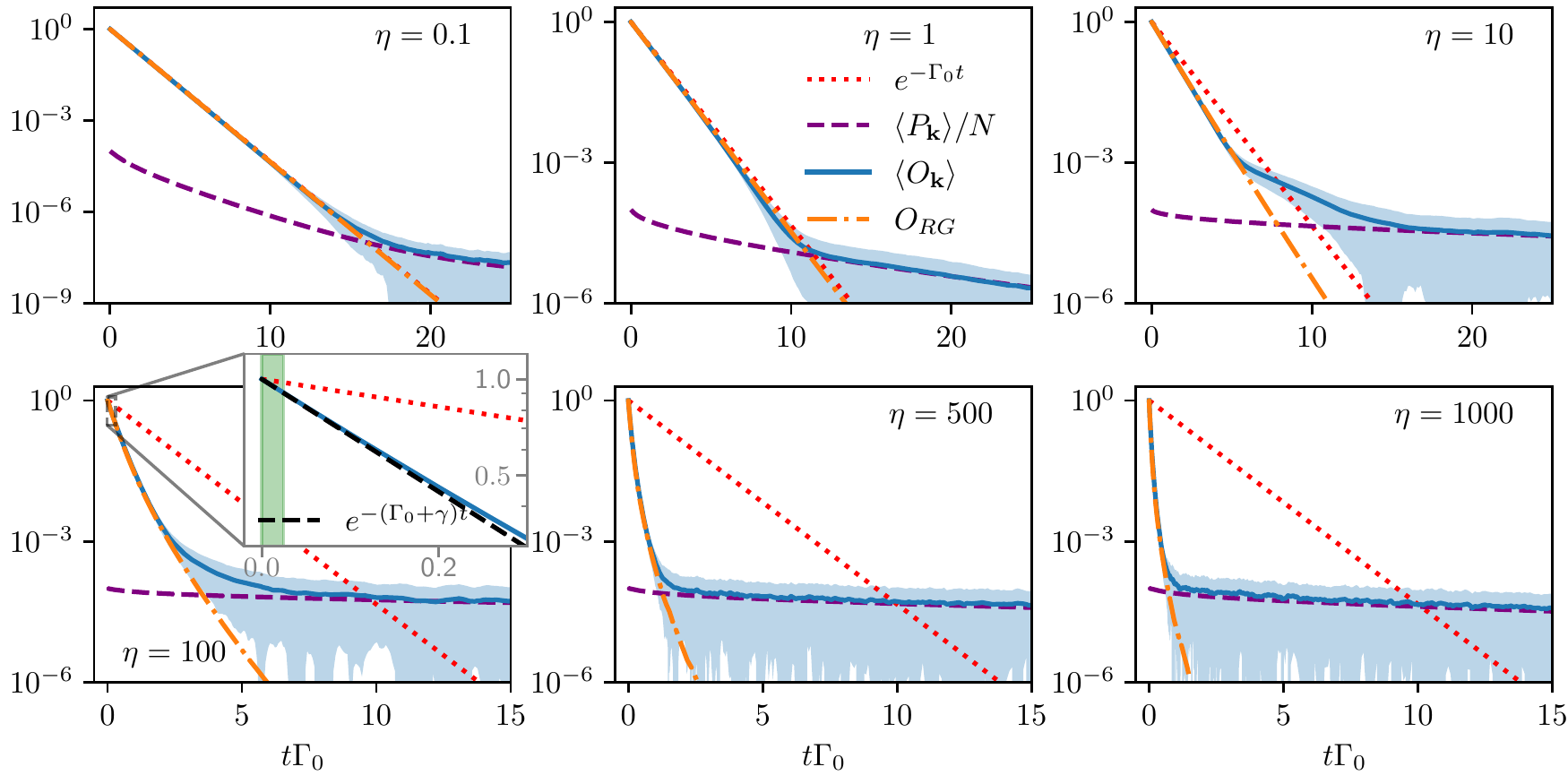}
    \caption{
    Time evolution of an initially prepared ideal mismatched spin-wave ($|\mathbf{k}|\neq k_0$), at different values of the dimensionless density, from small, $\eta=0.1$ to high $\eta=1000$.
    The blue line is the average time evolution of the overlap $O_{\mathbf{k}}$ (see Eq.\eqref{eq:Pop&Overlap}) over different realizations of the disordered gas, while the blue shaded region corresponds to the standard deviation.  
	The red dotted line shows an exponential decay with a rate $\Gamma_0$, as predicted by treating the atomic medium as smooth (MBE).
	The purple dashed line is the population of the time evolved state divided by the atom number, $P_{\mathbf{k}}/N$. Finally, the orange line is the overlap $O_{RG}$ as predicted by RG theory (see Sec.~\ref{sec:RG}). 
	We simulate the time evolution of $N=10^4$ atoms, to guarantee that, at the maximum density $\eta=1000$, the radius of the uniformly distributed spherical cloud is $R/\lambda_{eg}=1.34>1$, such that the cloud is not subwavelength. All the quantities are averaged over $N_s\sim 500$ different atomic samples. At density $\eta=100$, the inset shows the short-time exponential dynamics. The interval over which we fit for an exponential $e^{-(\Gamma_0+\gamma)t}$ (dashed black curve) is highlighted in green and corresponds to $\Gamma_{\mathbf{k}}t<0.1$, while at longer times the overlap deviates from this simple behavior, as predicted by our RG approach.}
    \label{fig:all_densities_N10000}
\end{figure*}
%------------------------------------------------------------%
We now present the exact numerical simulations and analysis of the time evolution of a mismatched spin wave for densities ranging from dilute ($\eta\ll 1$) to dense ($\eta\gg 1$), which we will then compare with the simple RG prediction of Eq.~(\ref{eq:overlap_RG}). To be concrete, we take an initial state consisting of a highly mismatched spin wave (Eq.~\eqref{eq:spin_wave}) with momentum $|\mathbf{k}|\!=\!6k_0$, $\hat{x}-$polarized and directed along $\hat{z}$, in an ensemble of $N\!=\!10^4$ atoms.
Then, the time evolution of the total excited-state population $P_{\mathbf{k}}(t)$ and of the overlap with the initially prepared spin-wave order $O_{\mathbf{k}}(t)$ is calculated for $N_s\!=\!500$ realizations of the disordered gas and averaged, as indicated by $\langle P_{\mathbf{k}}\rangle$ for example. Numerical results are represented in Fig.~\ref{fig:all_densities_N10000}, where we also plot our prediction for time evolution of the overlap, $O_{RG}$, made in Eq.\eqref{eq:overlap_RG} (orange dash-dotted lines), based on the effective single-atom theory described in the previous section.

\begin{figure}
    \centering
    \includegraphics[width=0.45\linewidth]{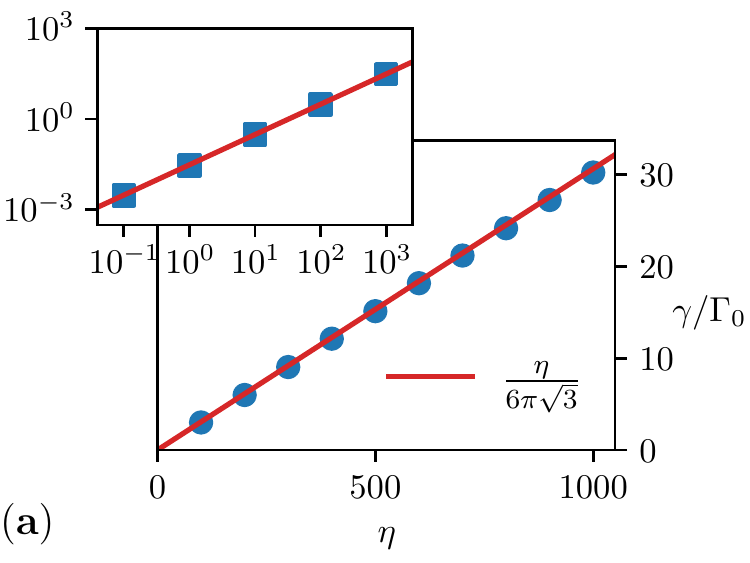}
    \includegraphics[width=0.45\linewidth]{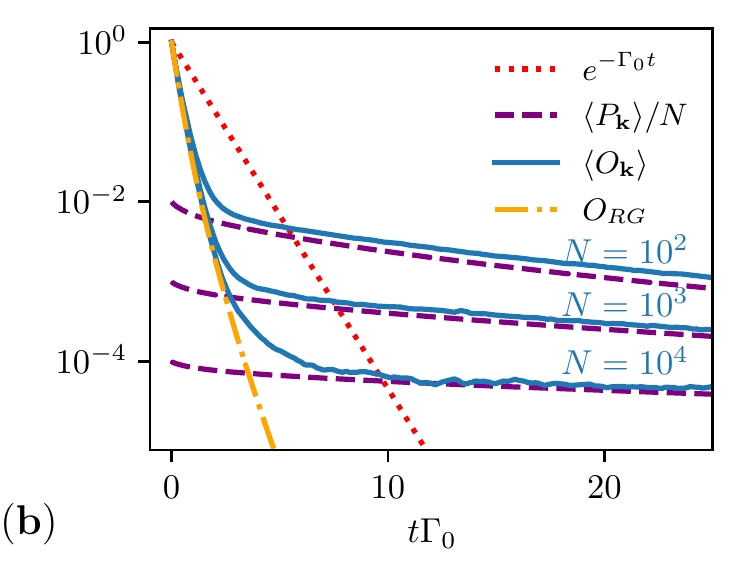}
        \caption{(a) Short-time dephasing rate vs. dimensionless density $\eta$. Blue dots are extracted from an exponential fit at short times ($\langle O_{\mathbf{k}}\rangle\sim e^{-\Gamma_{\mathbf{k}} \delta t}$), within the time interval $\Gamma_{\mathbf{k}}\delta t=0.1$, of the full time evolution of a mismatched spin-wave ($|\mathbf{k}|=6k_0$) at different densities $0.1 \leq \eta \leq 10^3$. The red line is instead the theoretical prediction for the short-time dephasing rate $\gamma=\Gamma_e\xi\eta$, discussed in the main text. Simulations are performed with a cloud of atoms of fixed size ($R/\lambda_{eg}=1.34$) and averaging over $N_s\!\sim\!500$ realizations. (inset) The same short time dephasing rate is evaluated (square dots), but now to explore the low-density regime we simulate the spin-wave dynamics in an ensemble with a fixed number of atoms, $N=10^4$ (again averaged $N_s\sim500$ times) and varying the radius of the cloud.
        (b) Numerical simulations of the time evolution of the averaged overlap $\langle O_{\mathbf{k}}\rangle$ (solid blue lines) for a mismatched spin-wave ($|\mathbf{k}|=6.0k_0$) compared with the RG prediction $O_{RG}$ (orange dash-dotted line), and the total excited population divided by atom number, $\langle P_k\rangle/N$ (purple dashed line). 
    The simulations consider a system of fixed density $\eta=100$, but different atom number $N=10^2,10^3,10^4$.
        }
    \label{fig:2l_short_time_Dephasing_and_finite_size}
\end{figure}

We first focus on the short-time dynamics.
As introduced in the previous section, the short-time decay of a spin-wave is predicted to be exponential \cite{He2021},  $\langle O_{\mathbf{k}}(\delta t)\rangle \sim e^{-\Gamma_{\mathbf{k}}\delta t}$, within a time interval $\Gamma_{\mathbf{k}}\delta t\ll 1$.
The rate $\Gamma_{\mathbf{k}}=\Gamma_0+\gamma$ is given by the sum of the single-atom emission rate, as predicted by the macroscopic MBE, and an additional density-dependent dephasing rate, $\gamma=\Gamma_0\xi\eta$, with $\xi=1/6\pi\sqrt{3}$. 
As shown in the inset of $\eta=100$ (Fig.~\ref{fig:all_densities_N10000}), for example, the microscopic dynamics reveal a short-time decay of the spin wave order that is distinctly faster than the MBE prediction. The decay becomes even more evident at higher densities~($\eta=10^2,10^3$). We can confirm the density dependence in the short-time decay rate $\Gamma_{\mathbf{k}}$ by fitting the curve $\langle O_{\mathbf{k}}\rangle$ for each density in the short-time window, defined by $\Gamma_\mathbf{k}\delta t<0.1$, to an exponential of the form $e^{-(\Gamma_0+\gamma)t}$, and plotting the dependence of the fit parameter $\gamma$ versus density parameter $\eta$ in Fig.~\ref{fig:2l_short_time_Dephasing_and_finite_size} (a). We also plot the prediction $\gamma=\Gamma_0\xi\eta$ in red. It is important to observe that within our defined ``short time'' interval, already $\sim 10\%$ of the initial spin-wave order is lost. An excellent agreement is observed over a large range of densities, both changing size and number of atoms (respectively blue squares and circles) of a random gas, confirming that this effect does only depend on the dimensionless density parameter $\eta$. This dephasing for moderate $\eta\lesssim 1$ was also observed experimentally in Ref.~\cite{He2021}.

%----------------------------------------------------------%
%----------------------------------------------------------%

At longer times and at higher densities, the decay of spin wave order noticeably deviates from exponential. Despite its simplicity, our single effective atom model based on RG (Eq.~\eqref{eq:overlap_RG}), displays excellent agreement beyond the short-time interval.  Viewed from the RG perspective, this  non-exponential contribution comes from the frequency components near the center of the inhomogeneous broadening probability distribution $P(\omega/\eta)$~(Fig.~\ref{fig:RG_Pw}), corresponding to atoms that are renormalized multiple times. Interestingly, at even longer times, it can be seen that for each density, the average  $\langle O_{\mathbf{k}}\rangle$ deviates from our prediction and saturates to a value that barely decreases over the range of times plotted. We furthermore observe numerically that this value closely coincides with the total excited state population remaining divided by the atom number, $P_{\bf k}(t)/N$~(dashed purple curves).

The slow decay of population $P_{\mathbf{k}}(t)$ at long times is an effect that has been studied extensively in recent years, and is known as late-time subradiance \cite{Guerin2016,Araujo2018,Weiss2018,Fofanov2020,Ferioli2021}. While a microscopic derivation is difficult, a heuristic argument can be made that the remaining population should be roughly equally distributed throughout the ensemble, given a smooth initial distribution. Furthermore, given the randomness of the dynamics, this population will be statistically evenly distributed among any N extended modes that can be defined for the system, such as our spin wave mode of interest.

As far as we can numerically check (e.g., up to $N=10^4$ atoms for a density of $\eta=100$ in Fig.~\ref{fig:2l_short_time_Dephasing_and_finite_size} (b)), we see that the RG prediction $O_{RG}$ follows the actual spin wave survival ratio $O_{\mathbf{k}}$ for increasingly long times as $N$ is increased, due to the decrease in the saturation value $P_\mathbf{k}/N$. This strongly suggests that the single-atom RG prediction should be interpreted as the correct description of the dephasing dynamics in the thermodynamic limit, when the late-time population in any one mode $\sim 1/N$ becomes negligible.

\section{\label{sec:conclusions}Conclusions}
In summary, we have developed an effective single-atom theory that describes well the non-exponential dephasing dynamics of optical spin waves in disordered atomic media, including at high densities and at long times. This theory is based upon the technique of strong disorder renormalization group, which treats the potentially strong near field interactions in such a medium in a non-perturbative way.

We envision that our predictions, particularly in the high density regime, could be immediately explored using solid-state emitter ensembles such as rare earth doped crystals \cite{DeRiedmatten2008,Hedges2010,Zhong2015,Zhong2015a}, where many atoms per cubic wavelength are typical. Separately, the remarkable accuracy by which RG is found to reproduce the dephasing dynamics suggests that it can be a powerful tool to quantitatively investigate and understand other microscopic optical phenomena in disordered systems.
\section{Acknowledgements}
We acknowledge F. Andreoli for stimulating discussions. DEC acknowledges support from the European Union’s Horizon 2020 research and innovation programme, under European Research Council grant agreement No 101002107 (NEWSPIN), FET-Open grant agreement No 899275 (DAALI), and Quantum Flagship project 820445 (QIA); the Government of Spain (Europa Excelencia program EUR2020-112155, Severo Ochoa program CEX2019-000910-S, and MICINN Plan Nacional Grant PGC2018-096844-B-I00); Generalitat de Catalunya through the CERCA program, Fundació Privada Cellex, AGAUR Project No. 2017-SGR-1334, Fundació Mir-Puig, and Secretaria d'Universitats i Recerca del Departament d'Empresa i Coneixement de la Generalitat de Catalunya, co-funded by the European Union Regional Development Fund within the ERDF Operational Program of Catalunya (project QuantumCat, ref. 001-P-001644).
YH and SW acknowledge support from National Key Research Program of China under Grant No.~2017YFA0304204, and from NSFC under Grant No.~12074083. 
%-----------------------------------%

\section*{References}
%\bibliographystyle{iopart-num}
%\bibliography{references}
%\bibliography{library}

\bibliographystyle{aip.bst}
\bibliography{library}

\end{document}